\documentclass[sigconf]{acmart}

\newcommand\Tstrut{\rule{0pt}{2.6ex}}         % = `top' strut
\newcommand\Bstrut{\rule[-0.9ex]{0pt}{0pt}}   % = `bottom' strut

\AtBeginDocument{%
  \providecommand\BibTeX{{%
    \normalfont B\kern-0.5em{\scshape i\kern-0.25em b}\kern-0.8em\TeX}}}

\copyrightyear{2024}
\acmYear{2024}
\setcopyright{acmlicensed}\acmConference[CIKM '24]{Proceedings of the 33rd ACM International Conference on Information and Knowledge Management}{October 21--25, 2024}{Boise, ID, USA}
\acmBooktitle{Proceedings of the 33rd ACM International Conference on Information and Knowledge Management (CIKM '24), October 21--25, 2024, Boise, ID, USA}
\acmDOI{10.1145/3627673.3679950}
\acmISBN{979-8-4007-0436-9/24/10}

\begin{document}

%% Original: Centralities of Nodes and Influence of Layers in Multiplex Networks
\title[LayerPlexRank: Exploring Node Centrality and Layer Influence \\ through Algebraic Connectivity in Multiplex Networks]{LayerPlexRank: Exploring Node Centrality and Layer Influence through Algebraic Connectivity in Multiplex Networks}

\author{Hao Ren}
\email{hao.ren@student.unsw.edu.au}
\orcid{0000-0003-2169-0111}
\affiliation{%
  \institution{University of New South Wales}
  \city{Sydney}
  \state{NSW}
  \country{Australia}
  \postcode{2052}
}

\author{Jiaojiao Jiang}
\authornote{Corresponding author.}
\email{jiaojiao.jiang@unsw.edu.au}
\orcid{0000-0001-7307-8114}
\affiliation{%
  \institution{University of New South Wales}
  \city{Sydney}
  \state{NSW}
  \country{Australia}
  \postcode{2052}
}

% Only used to shorten citations with three or more authors.
%\renewcommand{\shortauthors}{Ren et al.}
%\renewcommand{\shorttitle}{Calculation of Node Centrality and Layer Influence with Multiplex Network's Connectivity}

\begin{abstract}
  As the calculation of centrality in complex networks becomes increasingly vital across technological, biological, and social systems, precise and scalable ranking methods are essential for understanding these networks. This paper introduces LayerPlexRank, an algorithm that simultaneously assesses node centrality and layer influence in multiplex networks using algebraic connectivity metrics. This method enhances the robustness of the ranking algorithm by effectively assessing structural changes across layers using random walk, considering the overall connectivity of the graph. We substantiate the utility of LayerPlexRank with theoretical analyses and empirical validations on varied real-world datasets, contrasting it with established centrality measures.
\end{abstract}

\begin{CCSXML}
<ccs2012>
   <concept>
       <concept_id>10002950.10003624.10003633.10010917</concept_id>
       <concept_desc>Mathematics of computing~Graph algorithms</concept_desc>
       <concept_significance>500</concept_significance>
       </concept>
   <concept>
       <concept_id>10002950.10003624.10003633.10003644</concept_id>
       <concept_desc>Mathematics of computing~Network flows</concept_desc>
       <concept_significance>500</concept_significance>
       </concept>
   <concept>
       <concept_id>10002951.10003260.10003261.10003267</concept_id>
       <concept_desc>Information systems~Content ranking</concept_desc>
       <concept_significance>500</concept_significance>
       </concept>
   <concept>
       <concept_id>10002951.10003317.10003338.10003343</concept_id>
       <concept_desc>Information systems~Learning to rank</concept_desc>
       <concept_significance>300</concept_significance>
       </concept>
   <concept>
       <concept_id>10002951.10003317.10003338.10003342</concept_id>
       <concept_desc>Information systems~Similarity measures</concept_desc>
       <concept_significance>100</concept_significance>
       </concept>
   <concept>
       <concept_id>10003752.10003809.10003635.10003644</concept_id>
       <concept_desc>Theory of computation~Network flows</concept_desc>
       <concept_significance>300</concept_significance>
       </concept>
 </ccs2012>
\end{CCSXML}

\ccsdesc[500]{Mathematics of computing~Graph algorithms}
\ccsdesc[500]{Mathematics of computing~Network flows}
\ccsdesc[500]{Information systems~Content ranking}
\ccsdesc[300]{Information systems~Learning to rank}
\ccsdesc[100]{Information systems~Similarity measures}
\ccsdesc[300]{Theory of computation~Network flows}

\keywords{Centrality; Influence; Algebraic Connectivity; Multiplex Networks}

%\received{10 May 2024}
%\received[revised]{16 July 2024}
%\received[accepted]{8 August 2024}

\maketitle

%---------------------------------------------------------------------------------------------------
\section{Introduction}\label{sec:c1_intro}

The study of complex networks is pivotal in diverse fields, from sociology to computational biology, as it provides insights into the intricate web of real-world relationships \cite{gallotti2014anatomy, kinsley2020multilayer, coscia2013you}. A key focus within these studies is the centrality of nodes, which signifies their relative importance within the network \cite{lee2015towards}. In addressing node centrality, both single-layer (monoplex) and multi-layer (multiplex) networks provide crucial frameworks for analysis.

In monoplex networks, represented by the graph $G = (V, E)$ where $V$ denotes nodes and $E$ their connections, several centrality measures analyze node significance. Degree centrality gauges a node’s direct connections \cite{freeman1979centrality}, while betweenness centrality identifies nodes critical to network communication by their placement on shortest paths between others \cite{freeman1979centrality}. Closeness centrality measures the swiftness of information spread across the network from a node \cite{freeman1979centrality}, and eigenvector centrality assesses influence based on connectivity to other prominent nodes \cite{bonacich1972technique}.

Advancements like PageRank \cite{page1999pagerank} and HITS \cite{kleinberg1999authoritative} further enhance network analysis tools. PageRank evaluates node significance recursively through its connections, while HITS distinguishes nodes as hubs or authorities based on their outgoing or incoming connections, respectively. These methods have evolved to tackle complex network behaviors \cite{ng2001stable, ding2003pagerank}.

However, with the increasing complexity and layering of real-world networks \cite{Horvat2018}, single-layer network models often fall short. Real systems are often not isolated, and changes in information within a single level will have corresponding effects on other levels \cite{cosnet2024}. Multiplex networks, where multiple types of relationships coexist, are modeled as $G = (V, \bigcup_{\alpha=1}^M E^{[\alpha]})$ to reflect these complexities \cite{cozzo2013contact}. In such networks, centrality measures must consider both intra-layer and inter-layer connections to accurately gauge node significance.

An innovative approach in multiplex network analysis is the adaptation of traditional random walk centrality to effectively navigate layered structures \cite{wang2017identifying, sole2016random}. This measure evaluates the probability of a node being visited in a stochastic process across layers, providing a comprehensive view of its network role.

Our proposed algorithm, \textit{LayerPlexRank}, incorporates algebraic connectivity \cite{fiedler1973algebraic}—a metric used to assess cohesion and resilience in single-layer networks. By applying this metric to multiplex networks, LayerPlexRank determines how changes in one layer impact node centrality across the network, enhancing our understanding of network stability and interconnectedness. It computes node centralities and layer influences simultaneously, improving efficiency and allowing for dynamic adjustments in response to structural changes. This capability enables LayerPlexRank to adapt quickly to network variations and accurately reflect the dynamics of multiplex networks.

This paper discusses LayerPlexRank's theoretical foundations and its effectiveness through detailed simulations and empirical studies. The algorithm's integration of algebraic connectivity into multiplex centrality analysis provides new insights into the dynamics of complex, interconnected systems. The code implementation is available at \url{https://github.com/ninn-kou/LayerPlexRank}.

%---------------------------------------------------------------------------------------------------
\section{Methodology}\label{sec:c2_methods}

Assume the multiplex network comprises $M$ layers, each containing $N$ nodes. We denote the adjacency matrix of layer $\alpha$, where $1 \leq \alpha \leq M$, as $\mathbf{A}^{[\alpha]} = \left( A_{ij}^{[\alpha]} \right)_{1 \leq i, j \leq N}$. Additionally, $x_i$ represents the centrality of node $i$, and $z^{[\alpha]}$ indicates the influence of layer $\alpha$. After we merge all the layers into a heterogeneous network, we have
\begin{equation}\label{eq:w}
  W^{[\alpha]} = \sum_{i=1}^N\sum_{j=1}^N A_{ij}^{[\alpha]}.
\end{equation}

Additionally, according to Bianconi et al. \cite{rahmede2018centralities}, we leverage the weighted bipartite network, which is derived from the multiplex network and incorporates both nodes and layers \cite{boccaletti2014structure, cellai2016multiplex} as
\begin{equation}\label{eq:b}
  B_{i}^{[\alpha]} = \frac{\sum_{j=1}^N A_{ji}^{[\alpha]}}{W^{[\alpha]}}.
\end{equation}
This bipartite structure provides essential insights into the activity levels of nodes across each layer, revealing their presence and active participation within specific layers \cite{nicosia2015measuring}.

We employ the concept of random walks, where the computation of node centrality is approached as determining the stationary distribution of a Markov process. Nodes that are more frequently visited within this distribution have higher probability densities and thus are assigned greater centrality weights. We extend this concept to the centrality of layers, treating it akin to a one-dimensional Markov chain; thus, layers with higher weights exert more influence on the centrality calculations of nodes. After normalization, the aggregated probability density of a node across all layers indicates its overall importance, which then factors into subsequent computations. Additionally, we maintain the independence of nodes within individual layers to mitigate the influence of outliers, such as a single highly central node in an otherwise low-impact layer, on the overall centrality assessment. We set $\delta = 0.85$ as the damping factor, which represents the initial probability of a random walk moving to an adjacent node, analogous to the PageRank algorithm \cite{page1999pagerank}. The product of $(1 - \delta)$ and the proportion represented by $G_{ji}$ represents the likelihood of transitioning to a non-isolated node within $G$ \cite{rahmede2018centralities}.

Then, by given the influence of the layer $z^{[\alpha]}$, the centrality of node $i$ which denote as $x_i$ can be calculated as
\begin{equation}\label{eq:x}
  x_i = \sum_{\alpha=1}^M \sum_{j=1}^N \frac{B_{j}^{[\alpha]}  x_j}{\sum_{k=1}^N B_{k}^{[\alpha]}} + \frac{h_i}{\sum_{i=1}^N h_i} \sum_{j=1}^N (1-\delta \sum_{i=1}^N G_{ji}) x_j,
\end{equation}
where $\mathbf{G}$ denotes the weighted adjacency matrix of multiplex network $\mathbf{A}$ as
\begin{equation}
  G_{ij} = \sum_{\alpha=1}^M A_{ij}^{[\alpha]} z^{[\alpha]},
\end{equation}
and
\begin{equation}
  h_i = \sum_{j=1}^N (G_{ij} + G_{ji}).
\end{equation}

Furthermore, as a critical metric of network resilience, we consider the algebraic connectivity of each layer. Layers exhibiting stronger connectivity typically demonstrate enhanced robustness and improved synchronization capabilities \cite{jamakovic2008robustness}. Mathematically, the algebraic connectivity of a graph could be represent as the second-smallest eigenvalue of the Laplacian matrix \cite{fiedler1989laplacian}. For undirected graph, we have a symmetric and non-negative adjacency matrix that we use the symmetric normalized Laplacian \cite{butler2016algebraic} for layer $\alpha$ written as
\begin{equation}
  \mathcal{L}^{[\alpha]} = \mathbf{I} - \left( \mathbf{D^{[\alpha]}} \right) ^{-\frac{1}{2}} \mathbf{A}^{[\alpha]} \left( \mathbf{D^{[\alpha]}} \right) ^{-\frac{1}{2}}
\end{equation}
where $\mathbf{D^{[\alpha]}}$ is the degree matrix of layer $\alpha$. We use $\lambda_{2}(\mathcal{L}^{[\alpha]})$ to denote its second smallest eigenvalue, which is the algebraic connectivity.

In the calculation of layer influence, our algorithm draws partial inspiration from Ref. \cite{rahmede2018centralities}, particularly in recognizing that layers with highly central nodes are more influential. However, we have innovated and improved upon the previously mentioned methods. In determining node centrality, we consider characteristics of nodes that are layer-independent, and simultaneously, utilize algebraic connectivity to augment layer information in subsequent calculations of layer influence. This modification mitigates the impact of outliers, thereby enhancing the robustness and scalability of our algorithm. The influence of layer $\alpha$ denoted as $z^{[\alpha]}$ could be calculated as
\begin{align}
z^{[\alpha]} &= \frac{1}{N} \left[ \eta\lambda_{2}(\mathcal{L}^{[\alpha]}) + (1 - \eta) \left( W^{[\alpha]} \right)^a \left( \sum_{i=1}^N B_{i}^{[\alpha]} x_i^{s\gamma} \right)^s \right],
\end{align}
where $\eta$ is a parameter that controls the proportion of algebraic connectivity. We initially set it to $0.5$ to complete subsequent experiments.

Finally, similar to the approach in Ref. \cite{rahmede2018centralities}, the LayerPlexRank algorithm employs three parameters to adaptively assess layer and node influence based on network structure and node importance. These parameters are $a$, $s$, and $\gamma$, each with specific roles:
\begin{itemize}
  \item $a = 1$: The influence of a layer scales proportionally with its weight, $W^{\alpha}$.
  \item $a = 0$: The influence of a layer is normalized by its weight, $W^{\alpha}$, ensuring equal consideration irrespective of layer size.
  \item $s = 1$: Layers are considered more influential if they consist of nodes with higher centrality values.
  \item $s = -1$: Layers are deemed more influential if they are composed of fewer, but highly central nodes.
  \item $0 < \gamma < 1$: The influence of nodes with low centrality is either reduced ($s = -1$) or increased ($s = 1$) in their respective layers.
  \item $\gamma = 1$: The centrality of low centrality nodes is neither enhanced nor diminished.
  \item $\gamma > 1$: The influence of nodes with low centrality is either increased ($s = -1$) or reduced ($s = 1$), depending on their role in the layers.
\end{itemize}

%---------------------------------------------------------------------------------------------------
\section{Experiments and Evaluation}\label{sec:c3_expt}

To assess the performance of our algorithms on multiplex networks, we utilized four widely recognized real datasets to confirm their correctness, robustness and efficiency. In the experimental setup for LayerPlexRank, Unless otherwise noted, the values of $a$, $s$, and $\gamma$ were fixed at $1$ to ensure stable results.

\begin{figure*}
  \centering
  \includegraphics[scale=.26]{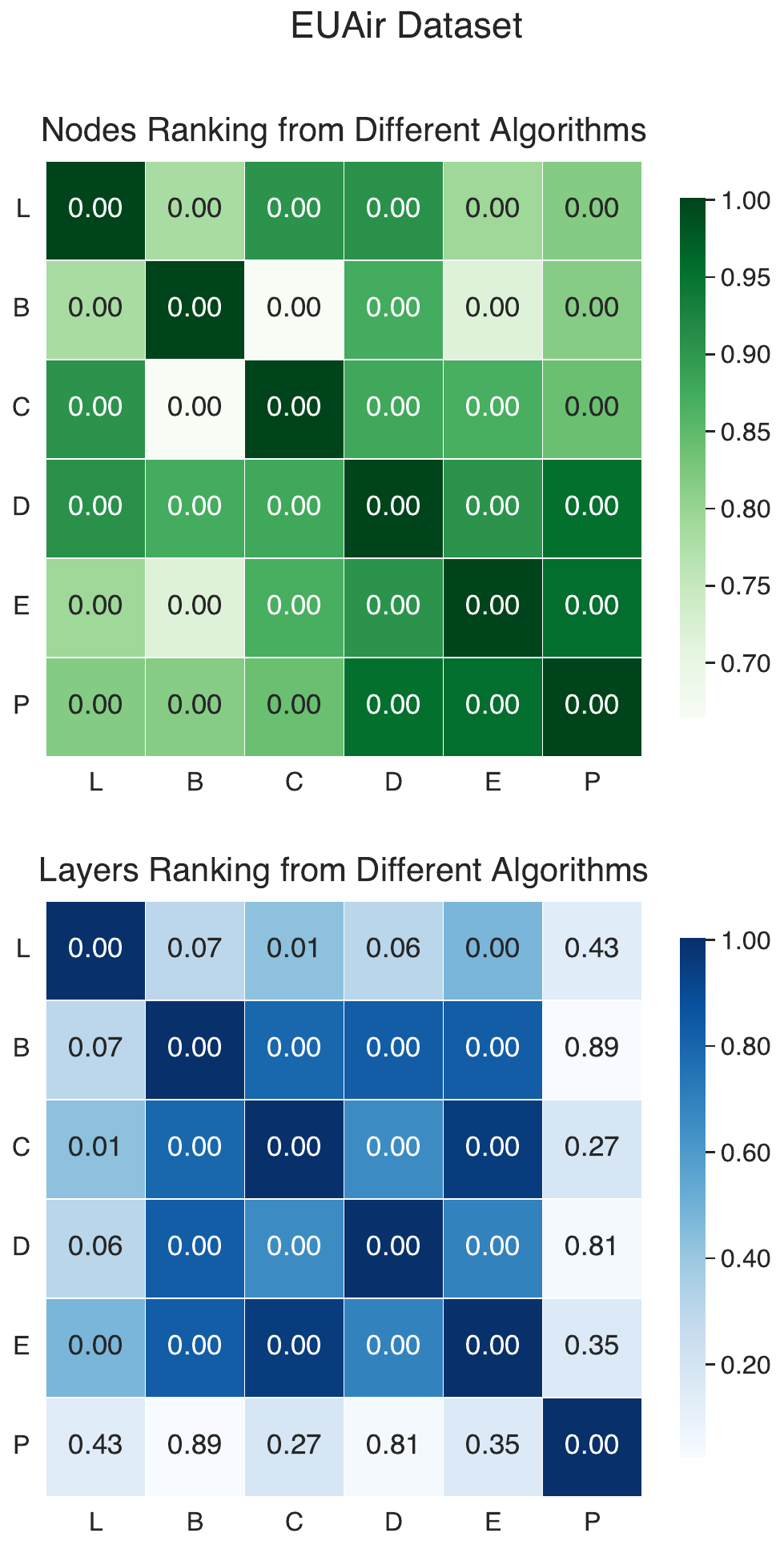}
  \includegraphics[scale=.26]{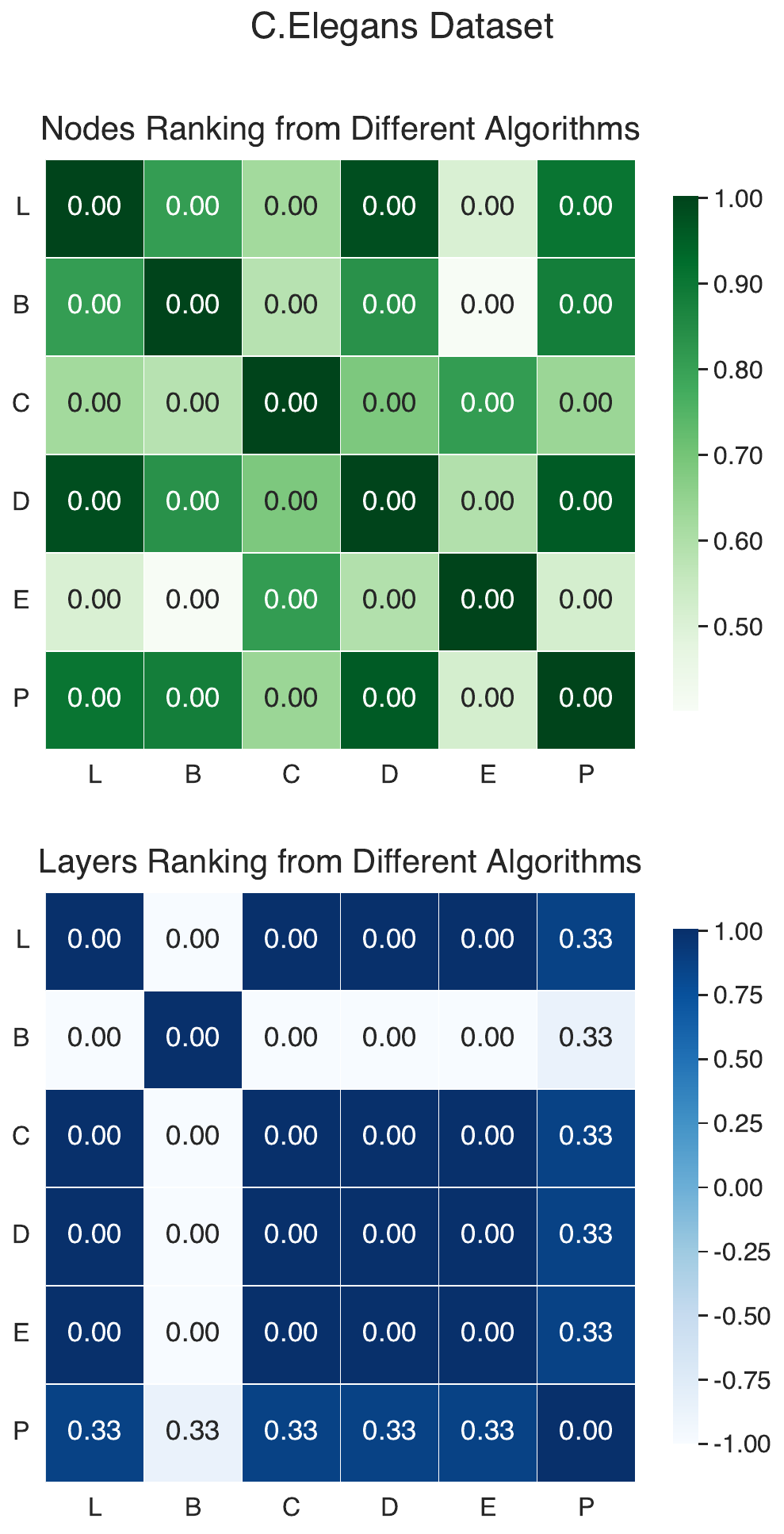}
  \includegraphics[scale=.26]{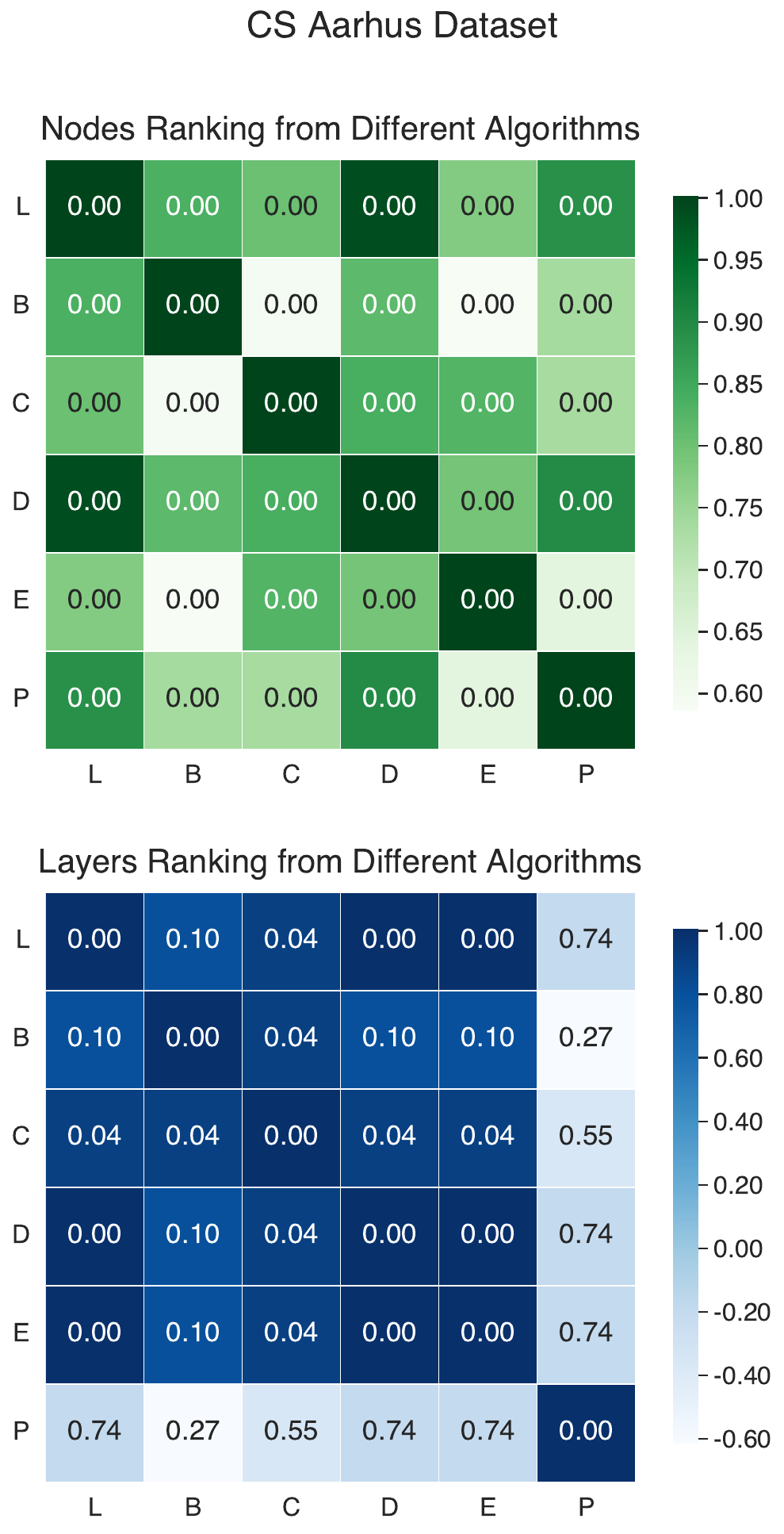}
  \includegraphics[scale=.26]{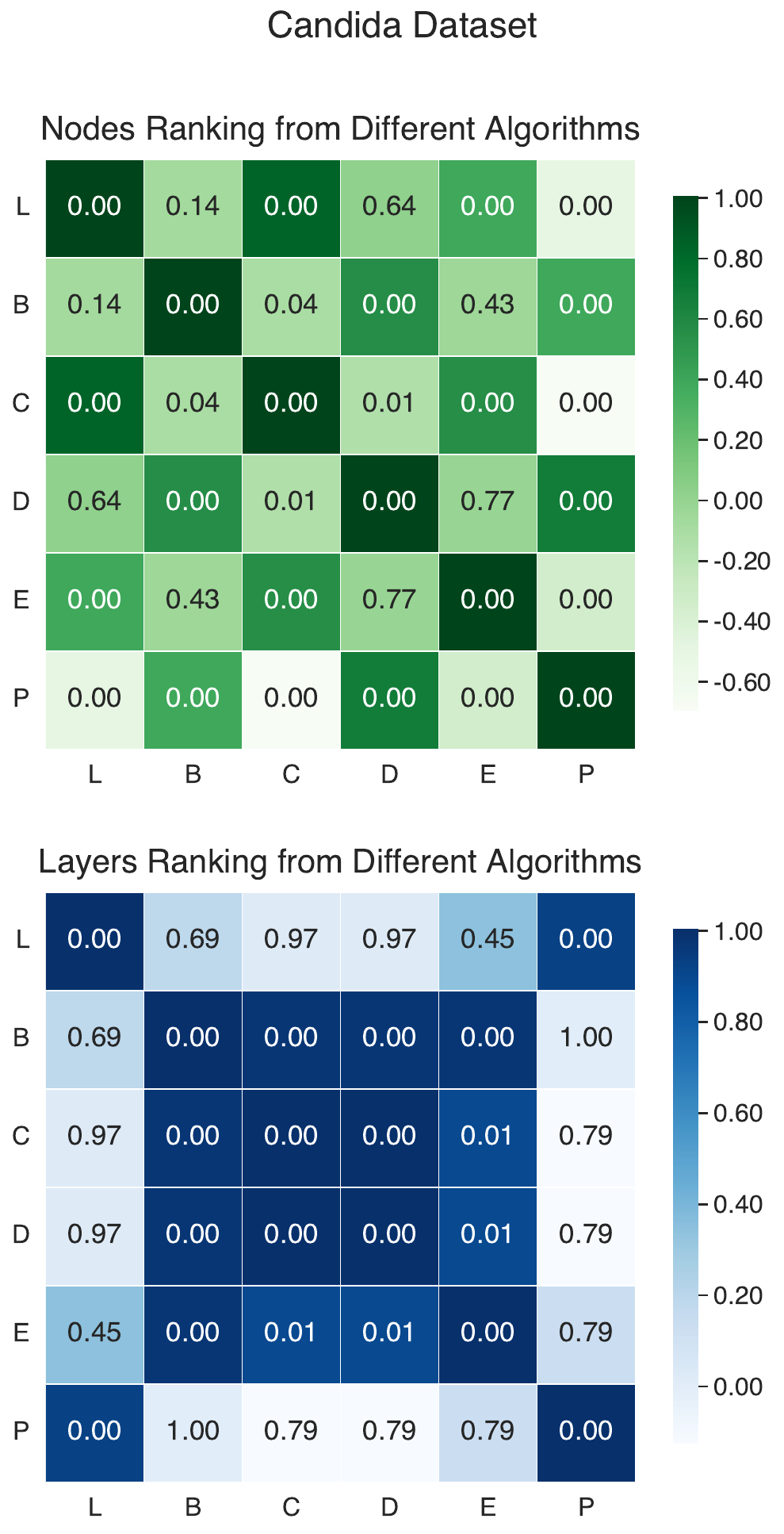}
  \caption{Spearman's Correlation matrix displays correlations between ranking results from algorithms (identified by their initials), with colour depth indicating correlation strength and numerical values showing confidence levels}
  \Description{A group of heatmaps was created using the Spearman correlation coefficient to represent the correlation between the results of various centrality calculation methods.}
  \label{fig:corr}
\end{figure*}

%---------------------------------------------------------------------------------------------------
\subsection{Datasets}

The \textbf{EUAir} dataset \cite{cardillo2013euair} features commercial flights between major European airports, forming a heterogeneous multiplex network with 37 airlines and 450 airports. The \textbf{CElegans} dataset \cite{beth2006celegans, demenico2014celegans} includes three layers of synaptic junctions: Electric, Chemical Monadic, and Chemical Polyadic. Additionally, \textbf{CS-Aarhus} \cite{magnani2013combinatorial} represents a social network with five types of relationships among Aarhus University's academic staff. The \textbf{Candida} dataset \cite{stark2006biogrid, de2015structural} from the BioGRID database details various genetic and protein interactions in Candida Albicans, including synthetic, suppressive, and additive genetic interactions; direct interactions; physical associations; general associations; and colocalizations. These last two datasets provide a basis for evaluating our method's performance on a smaller scale and with less established stability. Details of all datasets are summarized in Table \ref{tab:datasets}.

\begin{table}[h]
  \caption{Statistics of Used Multiplex Network Datasets}
  \label{tab:datasets}
  \begin{tabular}{c|c|c|c|c}
    \hline
    \textbf{Dataset} & \textbf{Layers} & \textbf{Nodes} & \textbf{Edges} & \textbf{Labels} \Tstrut\Bstrut\\
    \hline
    EUAir & 37 & 450 & 3588 & Undirected \Tstrut\Bstrut\\
    \hline
    CElegans & 3 & 279 & 5863 & Directed \Tstrut\Bstrut\\
    \hline
    CS-Aarhus & 5 & 61 & 620 & Undirected \Tstrut\Bstrut\\
    \hline
    Candida & 7 & 367 & 397 & Directed \Tstrut\Bstrut\\
    \hline
  \end{tabular}
\end{table}

%---------------------------------------------------------------------------------------------------
\subsection{Correctness Confirmation by Correlation with Baseline Methods}

Spearman's rank correlation coefficient \cite{spearman2010proof} can be used to measure the statistical dependence (i.e., the correlation between the ranking results of algorithms) as follows:
\begin{equation}
\rho = 1 - \frac{6 \sum \left( R(X_i) - R(Y_i) \right)^2}{n(n^2 - 1)},
\end{equation}
where $R(X)$ and $R(Y)$ represent two distinct rankings \cite{al2022toward} of identically sized samples $X$ and $Y$, respectively, and $n$ denotes the size of these samples.

To judge the correctness of the LayerPlexRank algorithm, we used Betweenness, Closeness, Degree, Eigenvector, and PageRank as benchmark centralities. According to Nicosia and Latora \cite{PhysRevE.92.032805}, there is an observed average correlation between the activity of nodes and their respective degrees within specified layers. Therefore, for monoplex-based algorithms, we calculated the layer influence by summing the centrality values of nodes within a single layer, and determined the total centrality for a specific node by aggregating its centrality values across all layers. A heatmap was then used to illustrate the correlations between the different ranking results from these algorithms, as depicted in Figure \ref{fig:corr}. The results clearly demonstrate that, although the rankings produced by LayerPlexRank and other commonly used algorithms may vary in detail due to different computational methods, the overall outcomes are highly interpretable. This indirectly validates the accuracy of the LayerPlexRank algorithm.

%---------------------------------------------------------------------------------------------------
\begin{figure*}
    \includegraphics[width=\linewidth]{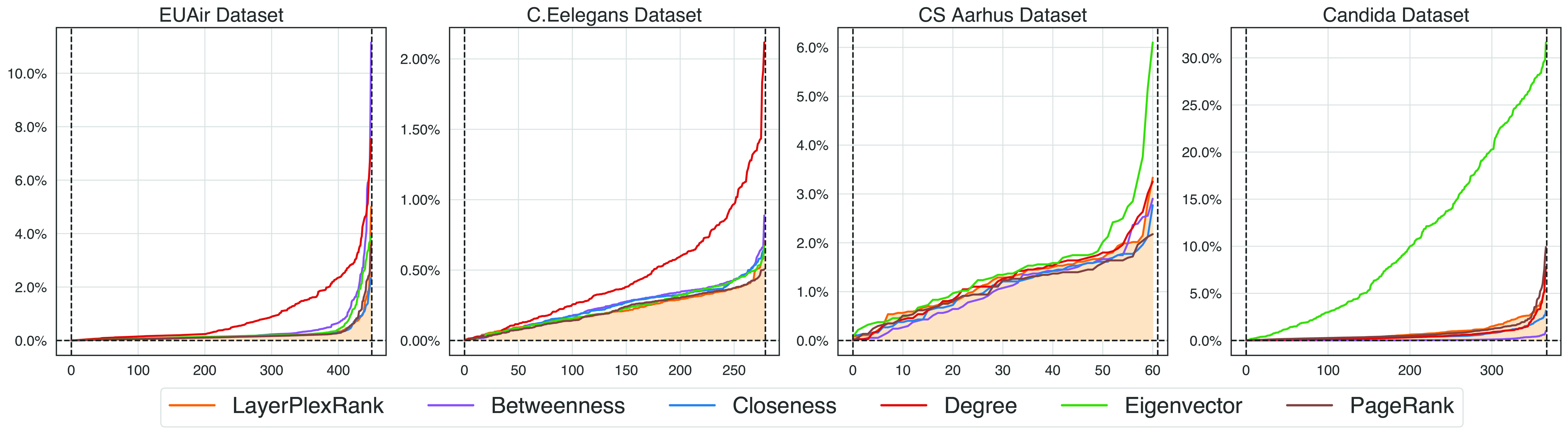}
    \caption{Algorithm's output versus LOOCV average, with nodes sorted by difference on the $x$-axis and values on the $y$-axis}
    \Description{Comparison of the algorithm's performance with the average results from the designed LOOCV, illustrating robustness. The area beneath each curve quantifies the aggregate differences, highlighting the stability of LayerPlexRank across multiple datasets.}
    \label{fig:diff}
\end{figure*}

\subsection{Robustness by Designed Cross-validation for Multiplex Networks}

In experiments, we demonstrate robustness using a specifically designed Leave-one-out Cross Validation (LOOCV) method \cite{sammut2011loocv}. LOOCV can be considered a specific application of cross-validation, which splits the original $n$-node dataset into an $n-1$ node training dataset and a $1$-node validation dataset in each iteration. The total number of validation times is $n$, indicating that all nodes have been left out once and used as a validation dataset. However, LOOCV focuses on multiple evaluations of the remaining training set using the split validation set, while our algorithm does not predict unknown data from existing data. Therefore, we use Jackknife resampling \cite{berger2006jackknife}, which omits a node and its edges in each iteration, complementing LOOCV to robustly aggregate results. This method highlights our algorithm's resilience to disruptions.

For each node $i$ (e.g., an airport in the \textit{EUAir} dataset), we remove it and its connections to form a new network, compute centrality measures, and derive a ranking list $R_i$ for the nodes in this altered graph. Moreover, we calculate the difference percentages between the average LOOCV results and one-time results for each node, as shown in Figure \ref{fig:diff}. The area below the curve represents the aggregate differences resulting from applying LOOCV to the algorithms. Compared with baseline methods, a smaller area under the curve for LayerPlexRank indicates greater robustness, demonstrated by stability and resilience against attacks involving node removal \cite{ma2023influence}.

%---------------------------------------------------------------------------------------------------
\subsection{Parameter Sensitivity Analysis}

\begin{figure}
    \includegraphics[scale=.31]{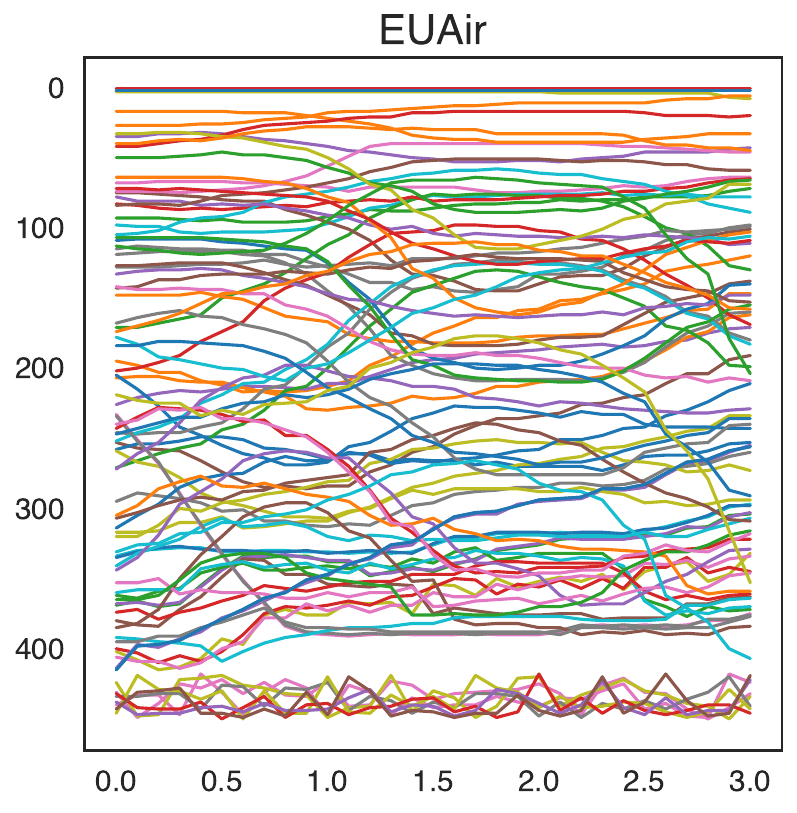}
    \includegraphics[scale=.31]{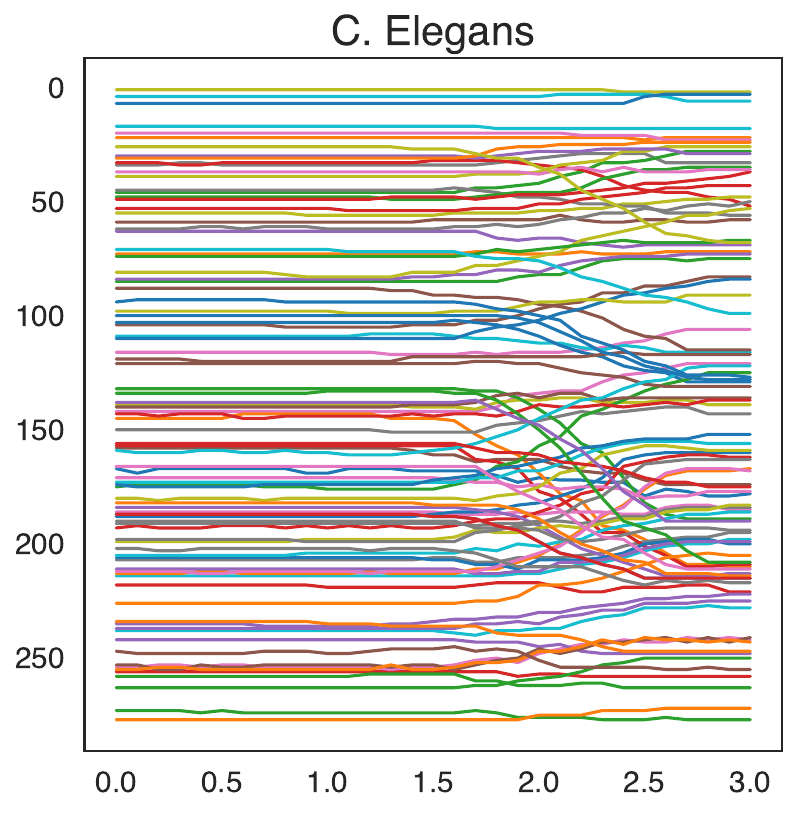}
    \includegraphics[scale=.31]{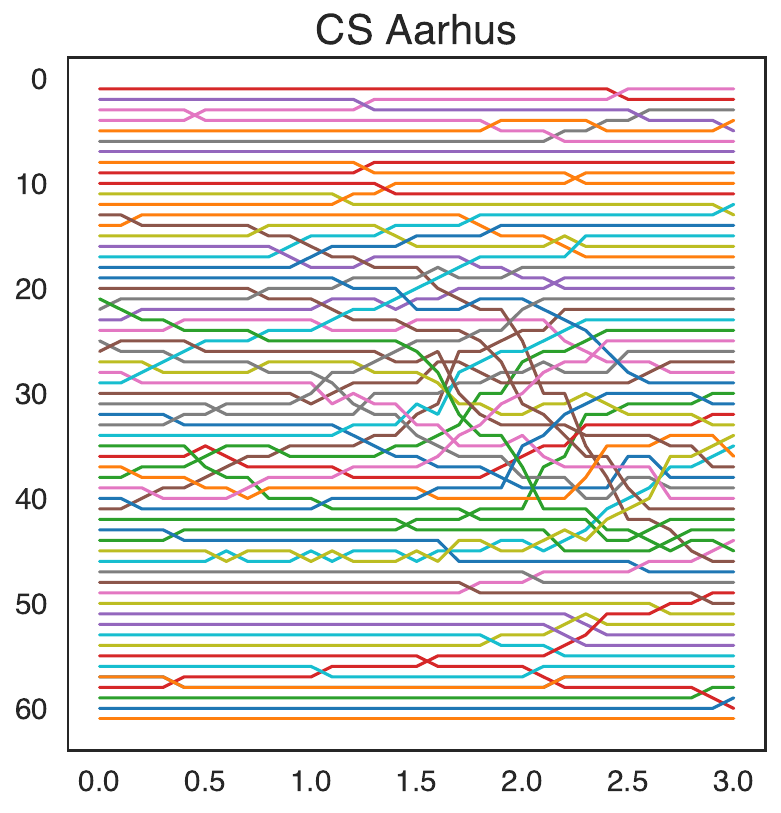}
    \includegraphics[scale=.31]{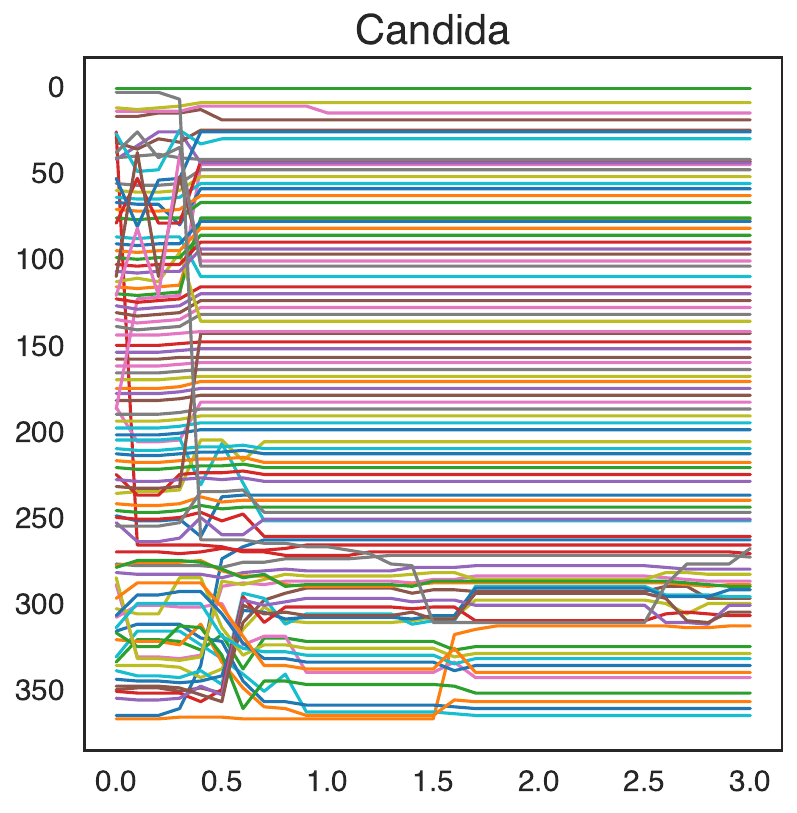}
    \caption{Parameter sensitivity analysis with $a = 1$ and $s = 1$, varying $\gamma = 0.1k$ within the range $k \leq 30, k \in \mathbb{N^+}$}
    \Description{Parameter sensitivity analysis highlighting the impact of varying gamma on node centrality across four datasets. Nodes are sampled uniformly for clarity, depicting the stability of highly central nodes and the general trend of centrality changes within the range gamma from 0 to 3.}
    \label{fig:sens}
\end{figure}

To evaluate the model's stability amid uncertainty, we conducted a sensitivity analysis on the parameter $\gamma$ \cite{razavi2015ofat}. This parameter dictates the contribution of nodes with relatively low centrality to the centrality calculation for this layer, where $\gamma < 1$ enhances and $\gamma > 1$ suppresses their influence.

For sensitivity testing, we set $a = 1$ and $s = 1$ to focus on the impact of other variables. We adjusted the value of $\gamma$ from $0$ to $3$, analyzing its effects across four datasets shown in Figure \ref{fig:sens}. While most nodes show a monotonic trend with varying $\gamma$, those with high centrality maintain stable rankings. To improve clarity and graph readability, we used a uniform sampling strategy, selecting every $\text{n}^{\text{th}}$ node to provide an unbiased and representative visualization of the trends.

%---------------------------------------------------------------------------------------------------
\section{Conclusions}\label{sec:c4_conclu}

LayerPlexRank advances multiplex network analysis by integrating node centrality and layer influence into a cohesive framework. Our algorithm effectively navigates the complexities of these networks, enhancing traditional methods with algebraic connectivity. This allows for the simultaneous consideration of multiple layers and their interactions. Extensive validation against established centrality measures and application to various real-world datasets demonstrate LayerPlexRank's ability to deliver nuanced insights into complex network structures and dynamics. Future refinement could expand centrality calculations to interconnected multilayered networks with explicit inter-layer connections, potentially pushing the boundaries of network science research.

\begin{acks}
  This work is supported by Australian Research Council Discovery Early Career Award (project number DE240101049).
\end{acks}

\bibliographystyle{ACM-Reference-Format}
\balance
\bibliography{references}

\end{document}